\documentclass[twocolumn,showpacs,preprintnumbers,amsmath,amssymb]{revtex4}
\usepackage{epsfig}
\usepackage{graphicx}
\usepackage{dcolumn}
\usepackage{bm}
\usepackage{amsthm}

\newcommand{\ket}[1]{\mbox{$ | #1 \rangle $}}
\newcommand{\bra}[1]{\mbox{$ \langle #1 | $}}

\begin{document}

\title{Sequential attacks against
differential-phase-shift quantum key distribution with weak
coherent states}

\author{Marcos Curty$^{1,2}$, Lucy Liuxuan Zhang$^{1}$,
Hoi-Kwong Lo$^{1}$, and Norbert L\"utkenhaus$^{2,3}$}
\affiliation{ 1 Center for Quantum Information and Quantum
Control, Department of Physics and Department of Electrical \&
Computer Engineering, University of Toronto, Toronto, Ontario,
M5S 3G4, Canada \\
2 Institute for Quantum Computing, University of Waterloo,
Waterloo, Ontario, N2L 3G1, Canada \\
3 Quantum Information Theory Group, Institut f\"ur Theoretische
Physik I, and Max-Planck Research Group, Institute of Optics,
Information and Photonics, Universit\"at Erlangen-N\"urnberg,
91058 Erlangen, Germany }

\date{\today}

\begin{abstract}
We investigate limitations imposed by sequential attacks on the
performance of differential-phase-shift quantum key distribution
protocols that use pulsed coherent light. In particular, we
analyze two sequential attacks based on unambiguous state
discrimination and minimum error discrimination, respectively, of
the signal states emitted by the source. Sequential attacks
represent a special type of intercept-resend attacks and,
therefore, they do not allow the distribution of a secret key.
\end{abstract}


\maketitle

\section{INTRODUCTION}

Quantum key distribution (QKD) \cite{gisin_rev_mod} is a
technique that exploits quantum effects to establish a secure
secret key between two parties (usually called Alice and Bob).
This secret key is the essential ingredient of the one-time-pad
or Vernam cipher \cite{vernam}, the only known encryption method
that can provide information-theoretic secure communications.

The first complete scheme for QKD is that introduced by Bennett
and Brassard in 1984 (BB84 for short) \cite{BB84}. A full proof
of the security for the whole protocol has been given in
Ref.~\cite{Mayers98}. After the first demonstration of the
feasibility of this scheme \cite{Bennett92}, several long-distance
implementations have been realized in the last years (see, for
instance, Ref.~\cite{Marand95} and references therein). However,
these practical approaches differ in many important aspects from
the original theoretical proposal, since that demands
technologies that are beyond our present experimental capability.
Specially, the signals emitted by the source, instead of being
single-photons, are usually weak coherent pulses (WCP) with
typical average photon numbers of $0.1$ or higher. Moreover, the
detectors employed by the receiver have a low detection
efficiency and are noisy due to dark counts. These facts,
together with the loss and the noise introduced by the quantum
channel, jeopardize the security of the protocol, and leads to
limitations of rate and distance that can be covered by these
techniques \cite{Huttner95,Norbert00}.

The main security threat of QKD based on WCP arises from the fact
that some pulses contain more than one photon prepared in the
same polarization state. Now, an eavesdropper (Eve) can perform,
for instance, the so-called {\it Photon Number Splitting} (PNS)
attack on the multi-photon pulses \cite{Huttner95}. This attack
provides Eve with full information about the part of the key
generated from the multi-photon signals, without causing any
disturbance in the signal polarization. As a result, it turns out
that the BB84 protocol with WCP can give a key generation rate of
order $O(\eta^2)$, where $\eta$ denotes the transmission
efficiency of only the quantum channel \cite{inamori}.

To obtain higher secure key rates over longer distances,
different QKD schemes robust against the PNS attack have been
proposed in recent years. One of these schemes is the so-called
decoy-states \cite{decoy_t,decoy_e}, where Alice randomly varies
the mean photon number of the signal states sent to Bob by using
different intensity settings. This technique delivers a key
generation rate of order $O(\eta)$ \cite{decoy_t,decoy_e}. Other
possibility is based on the transmission of two non-orthogonal
coherent states together with a strong reference pulse
\cite{ben92}. This scheme has been analyzed in
Ref.~\cite{koashi04}, where it was confirmed that also in this
scenario the secure key rate is of order $O(\eta)$. Finally,
another possible approach is the use of differential-phase-shift
(DPS) QKD protocols
\cite{dpsqkd,dpsqkd2,dpsqkd_exp1,dpsqkd_exp2}. In this kind of
schemes, Alice sends to Bob a train of WCP whose phases are
randomly modulated by $0$ or $\pi$. On the receiving side, Bob
measures out each incoming signal by means of an interferometer
whose path-length difference is set equal to the time difference
between two pulses. In this case, however, a secure key rate of
order $O(\eta)$ has only been proven so far against a particular
type of individual attacks where Eve acts on {\it photons}
individually, rather than {\it signals} \cite{dpsqkd2}. Whether
DPS QKD is secure against the most general attack remains an
important open question.

In this paper, we investigate limitations imposed by sequential
attacks on the performance of DPS QKD protocols. In this kind of
attacks, Eve measures out every coherent state emitted by Alice
and prepares new signal states, depending on the results obtained,
that are given to Bob. Whenever Eve obtains a predetermined
number of consecutive successful measurement outcomes, then she
prepares a train of WCP that is forwarded to Bob. Otherwise, Eve
sends vacuum signals to Bob to avoid errors. Sequential attacks
constitute a special type of intercept-resend attacks
\cite{jahma01,Felix01,curty05} and, therefore, they do not allow
the distribution of a secret key \cite{Curty04}. Here we shall
consider a conservative definition of security, {\it i.e.}, we
assume that Eve can control some flaws in Alice's and Bob's
devices ({\it e.g.}, the detection efficiency and the dark count
probability of the detectors), together with the losses in the
channel, and she exploits them to obtain maximal information
about the shared key.

We analyze two possible sequential attacks. In the first one, Eve
realizes unambiguous state discrimination (USD) of Alice's signal
states \cite{usd,chef,jahma01}. When Eve identifies unambiguously
a signal state sent by Alice, then she considers this result as
successful. Otherwise, she considers it a failure. In the second
attack, Eve performs first a filtering operation on each signal
emitted by Alice and, afterwards, she measures out each
successful filtered state following the approach of minimum error
discrimination (MED) \cite{hels,med2}, {\it i.e.}, she guesses the
identity of the filtered state with the minimum probability of
making an error. (See also Ref.~\cite{curty05}.) As a result, we
obtain upper bounds on the maximal distance achievable by DPS QKD
schemes as a function of the error rate in the sifted key, the
double click rate at Bob's side, and the mean photon-number of
the signals sent by Alice.

Instead of using an USD measurement on each signal state sent by
Alice, like in the first sequential attack that we consider, Eve
could as well employ the same detection device like Bob. This
sequential attack was very briefly introduced in
Ref.~\cite{dpsqkd2}. A successful result is now associated with
obtaining a click in Eve's apparatus, while a failure corresponds
to the absence of a click. However, since Alice's signal states
are typically coherent pulses with small average photon number,
the probability of obtaining a successful result in this scenario
is always smaller than the one of a sequential USD attack.
Therefore, a sequential USD attack can provide tighter upper
bounds on the performance of DPS QKD protocols than those derived
from an eavesdropping strategy where Eve uses the same
measurement apparatus like Bob.

A different QKD scheme, but also related to DPS QKD protocols,
has been proposed recently in Ref.~\cite{gisin1}. (See also
Ref.~\cite{gisin2}.) However, since the abstract signal structure
of this protocol is different from the one of DPS QKD schemes,
the analysis contained in this paper does not apply to that
scenario. Sequential attacks against the QKD protocol introduced
in Ref.~\cite{gisin1} have been investigated in
Ref.~\cite{valerio} following a similar approach like in this
paper.

The paper is organized as follows. In Sec.~\ref{sec_1} we
describe in more detail DPS QKD protocols. Then, in
Sec.~\ref{seqattack}, we present sequential attacks against DPS
QKD schemes. Sec.~\ref{susda} includes the analysis for a
sequential USD attack. Here we obtain an upper bound on the
maximal distance achievable by DPS QKD schemes as a function of
the error rate, the double click rate at Bob's side, and the mean
photon-number of Alice's signal states. Similar results are
derived in Sec.~\ref{medattack}, now for the case of sequential
attacks based on MED of the signals sent by Alice. Finally,
Sec.~\ref{CONC} concludes the paper with a summary.

\section{DIFFERENTIAL-PHASE-SHIFT QKD}\label{sec_1}

\begin{figure}
\begin{center}
\includegraphics[scale=1]{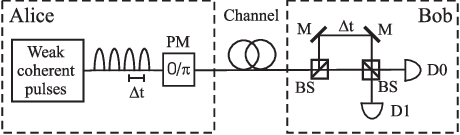}
\end{center}
\caption{Basic setup of a DPS QKD scheme. PM denotes a phase
modulator, BS, a $50:50$ beam-splitter, M, a mirror, D0 and D1 are
two photon detectors, and $\Delta{}t$ represents the time
difference between two consecutive pulses. \label{dpsqkd}}
\end{figure}

The setup is illustrated in Fig.~\ref{dpsqkd}
\cite{dpsqkd,dpsqkd2,dpsqkd_exp1,dpsqkd_exp2}.
Alice prepares first a train of coherent states $\ket{\alpha}$
and, afterwards, she modulates, at random and independently every
time, the phase of each pulse to be $0$ or $\pi$. As a result,
she produces a random train of signal states $\ket{\alpha}$ or
$\ket{-\alpha}$ that are sent to Bob through the quantum channel.
On the receiving side, Bob uses a $50:50$ beam-splitter to divide
the incoming pulses into two possible paths and then he recombines
then again using another $50:50$ beam-splitter. The time delay
introduced by Bob's interferometer is set equal to the time
difference $\Delta{}t$ between two pulses. Whenever the relative
phase between two consecutive pulses is $0$ ($\pm\pi$) only the
photon detector $D0$ ($D1$) may produce a ``click" (at least one
photon is detected). For each detected event, Bob records the
exact time where he obtained a click and the actual detector that
fired.

Once the quantum communication phase is completed, Bob uses a
classical authenticated channel to announce the time instances
where he detected at least one photon. From this information,
together with the knowledge of the phase value used to modulate
each pulse, Alice can infer which photon detector fired at Bob's
side each given time. Then, Alice and Bob can agree, for instance,
to select a bit value ``0" whenever the photon detector $D0$
clicked, and a bit value ``1" if the detector $D1$ fired. In an
ideal scenario, Alice and Bob end up with an identical string of
bits representing the {\it sifted key}. Due to the noise
introduced by the quantum channel together with possible
imperfections of Alice's and Bob's devises, however, the sifted
key typically contains some errors. Then, Alice and Bob perform
error-correction to reconcile the data, and privacy amplification
to decouple the data from Eve. (See, for instance,
Ref.~\cite{gisin_rev_mod}.)

In the next section we analyze simple sequential attacks against
the DPS QKD protocol introduced above that are particularly suited
for the signal states and detection methods employed by Alice and
Bob, together with the attenuation introduced by the channel. Let
us emphasize here that these attacks might not be optimal, but, as
we will show below, they already impose strong restrictions on
the performance of DPS QKD schemes with weak coherent pulses.

\section{Sequential attacks against DIFFERENTIAL-PHASE-SHIFT QKD}
\label{seqattack}

A sequential attack can be seen as a special type of
intercept-resend attack. First, Eve measures every coherent state
emitted by Alice with a detection apparatus located very close to
the sender. Afterwards, she transmits each measurement result
through a lossless classical channel to a source close to Bob.
Whenever Eve obtains a predetermined number of consecutive {\it
successful} measurement outcomes, this source prepares a train of
new signal states that is forwarded to Bob. Otherwise, Eve sends
vacuum signals to Bob to avoid errors. Whether a measurement is
considered to be successful or not and which type of non-vacuum
states Eve sends to Bob depends on Eve's particular eavesdropping
strategy and on her measurement device. Sequential attacks
transform the original quantum channel between Alice and Bob into
an entanglement breaking channel \cite{Horodecki03} and,
therefore, they do not allow the distribution of a secret key
\cite{Curty04}.

We begin by introducing Eve's measurement apparatus. As mentioned
previously, we shall consider two possible alternatives. Each
alternative provides a different sequential attack. In the first
one, Eve realizes USD \cite{usd,chef} of Alice's signal states.
Whenever Eve identifies unambiguously a signal state sent by
Alice, {\it i.e.}, she determines without error whether it is
$\ket{\alpha}$ or $\ket{-\alpha}$, she considers this result as
successful. If the measurement outcome corresponds to an
inconclusive result then she considers it a failure. The second
eavesdropping strategy can be decomposed into two steps: first,
Eve performs a filtering operation on each signal state sent by
Alice with the intention to make them, with some finite
probability, more ``distinguishable''. A failure refers now to
those signal states for which the filtering operation does not
succeed. Afterwards, Eve measures out each successful filtered
state following the approach of MED \cite{hels,med2}. Her goal is
to guess the identity of the filtered states with the minimum
probability of making an error. Notice that the first
eavesdropping strategy can be considered as a special case of the
second eavesdropping strategy where the probability that Eve
makes an error in distinguishing a state $\ket{\alpha}$ and
$\ket{-\alpha}$ is exactly zero. We shall denote as $p_{succ}$
the probability that Eve obtains a successful result whatever the
measurement device she employs.

\begin{figure}
\begin{center}
\includegraphics[scale=1]{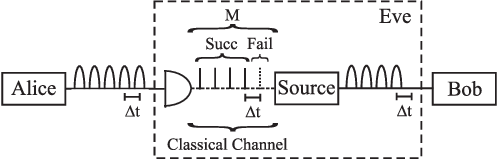}
\end{center}
\caption{Illustration of a sequential attack. In this example the
length of each block is $M=5$, the minimum number of consecutive
successful results within a block is given by $M_{min}=3$, and we
assume that Eve obtains $m=4$ consecutive successful results
within a block. A successful outcome is represented with a
vertical solid line in the classical channel, while a failure
result is denoted with a vertical dashed line. \label{dpsqkd2}}
\end{figure}
In order to evaluate her measurement outcomes, we shall consider
that Eve divides her data into different blocks of length $M$,
where each block contains $M$ consecutive measurement results.
Moreover, we assume that Eve analyzes each block of data
independently, {\it i.e.}, without considering the data included
in other blocks. As we will show later on, this eavesdropping
strategy will necessarily create some error rate that decreases
when incrementing the block length M. In this scenario, we define
the integer parameter $M_{min}$, with
$\lfloor{}M/2+1\rfloor\leq{}M_{min}<M$, as the minimum number of
consecutive successful results within a block that Eve needs to
obtain in order to send Bob a new train of coherent states
$\ket{\beta{}e^{i\theta_j}}$. This definition of $M_{min}$ arises
from the particular eavesdropping strategies that we consider
here, and the role of this parameter $M_{min}$ will become clear
later on. More precisely, if $m$ denotes the total number of
consecutive successful outcomes obtained by Eve within a block,
then, whenever $m$ is bigger than $M_{min}$, Eve prepares $m$
consecutive coherent states $\ket{\beta{}e^{i\theta_1}},
\ket{\beta{}e^{i\theta_2}},\ldots,\ket{\beta{}e^{i\theta_m}}$,
together with $M-m$ vacuum states for those unsuccessful results
within the block and sends these signals to Bob. On the other
hand, if $m<M_{min}$ Eve sends to Bob $M$ vacuum states. The case
$m=M_{min}$ deserves a special attention. We shall consider that
in this case Eve employs a probabilistic strategy that combines
the two previous ones. In particular, we assume that Eve sends to
Bob $M_{min}$ consecutive coherent states
$\ket{\beta{}e^{i\theta_1}},
\ket{\beta{}e^{i\theta_2}},\ldots,\ket{\beta{}e^{i\theta_{M_{min}}}}$
with probability $q$ and, with probability $1-q$, she sends to
Bob $M$ vacuum states. That is, the parameter $q$ allows Eve to
smoothly fit her eavesdropping strategy to the observed data.

The angle $\theta_j$ of a coherent state
$\ket{\beta{}e^{i\theta_j}}$ prepared by Eve depends on her
particular measurement strategy. When she utilizes an USD
measurement, then $\theta_j=0$ if the state identified by her
measurement is $\ket{\alpha}$, and $\theta_j=\pi$ if the state
identified is $\ket{-\alpha}$. A similar criterion can also be
applied to the case where Eve performs a filtering operation
followed by a MED measurement on the successful filtered states:
If the result obtained is associated with the signal state
$\ket{\alpha}$ then $\theta_j=0$, otherwise $\theta_j=\pi$.
Fig.~\ref{dpsqkd2} shows a graphical representation of such a
sequential attack for the case $M=5$ and $M_{min}=3$. In this
example, moreover, we assume that Eve obtains $m=4$ consecutive
successful results within a block.

Next, we obtain an expression for the Gain of a sequential attack,
{\it i.e.}, the probability that Bob obtains a click per signal
state sent by Alice, as a function of the parameters $M$,
$M_{min}$, $q$, the probability $p_{succ}$ of obtaining a
successful result, and the mean photon-number
$\mu_\beta=\vert\beta\vert^2$ of the coherent states sent by Eve.
Afterwards, we study the two sequential attacks introduced above
in more detail. The objective is to find an expression for the
quantum bit error rate (QBER) introduced by Eve, and for the
resulting double click rate at Bob's side in each of these two
attacks.

\subsection{Gain of a sequential attack}

The Gain of a sequential attack is defined as $N_{clicks}/N$,
where $N_{clicks}$ represents the average total number of clicks
obtained by Bob, and $N$ is the total number of signal states
sent by Alice. In this definition, we consider that double clicks
contribute to $N_{clicks}$ like single clicks. The parameter
$N_{clicks}$ can be expressed as $N_{clicks}=(N/M)N_{clicks}^M$,
with $N_{clicks}^M$ denoting the average total number of clicks
per block of length $M$ at Bob's side. With this notation, the
Gain of a sequential attack, that we shall denote as $G$, can then
be written as
\begin{equation}\label{gain}
G=\frac{1}{M}N_{clicks}^M.
\end{equation}

\begin{figure}
\begin{center}
\includegraphics[scale=0.9]{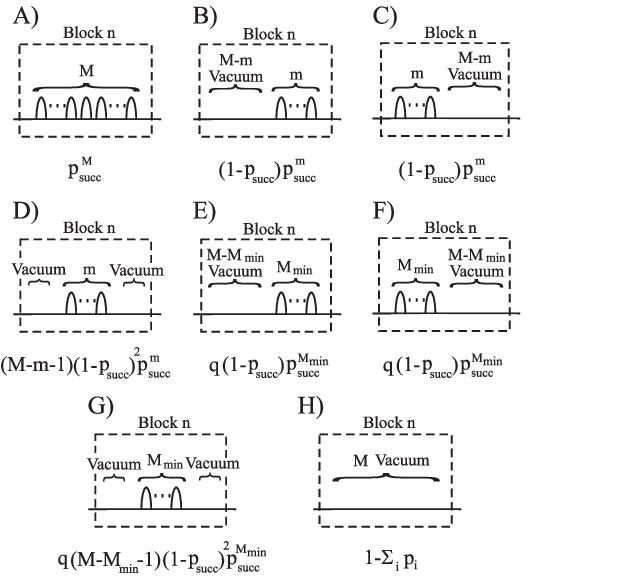}
\end{center}
\caption{Possible blocks of $M$ signals that Eve sends to Bob
together with their a priori probabilities. Case A: The block
contains $M$ coherent states. Case B: The first $m\in(M_{min},M)$
signals of the block are coherent states, while the last $M-m$
signals are vacuum states. Case C: The block contains first $M-m$
vacuum states followed by $m\in(M_{min},M)$ coherent states. Case
D: The block has $m\in(M_{min},M)$ coherent states and, at least,
the first and the last signal of the block are vacuum states.
Case E: The first $M_{min}$ signals of the block are coherent
states, while the last $M-M_{min}$ signals are vacuum states.
Case F: The block contains first $M-M_{min}$ vacuum states
together with $M_{min}$ coherent states. Case G: The block has
$M_{min}$ coherent states and, at least, the first and the last
signal of the block are vacuum states. Case H: The block contains
only vacuum states. The a priori probability of this last
scenario is given by $1-\sum_i p_i$, with $p_i$ representing the
a priori probabilities of each of the previous cases.
\label{alternatives}}
\end{figure}

\begin{figure}
\begin{center}
\includegraphics[scale=1]{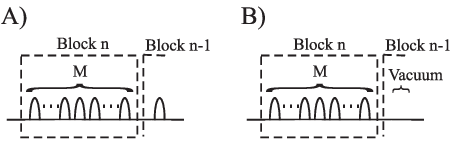}
\end{center}
\caption{Eve sends to Bob a block of signals containing M
coherent states (Block $n$ in the Figure). Case A: with
probability $p$, where $p$ is given by Eq.~(\ref{p_bar}), the
last signal state of the previous block is a coherent state. Case
B: with probability $1-p$ the last signal state of the previous
block is a vacuum state. \label{nueva_figura2}}
\end{figure}

Next, we obtain an expression for $N_{clicks}^M$. We shall
distinguish several cases, depending on the number $m$ of coherent
states $\ket{\beta{}e^{i\theta_1}},
\ket{\beta{}e^{i\theta_2}},\ldots,\ket{\beta{}e^{i\theta_m}}$ that
Eve sends to Bob inside a given block and the position of these
coherent states in the block \cite{note1}. These cases are
illustrated in Fig.~\ref{alternatives}, where we also include the
{\it a priori} probabilities to be in each of these scenarios.
Note, however, that the average total number of clicks in each of
these cases will also depend on whether the last signal state of
a previous block is actually a coherent state or not. To include
this boundary effect between blocks in our analysis, we shall
always distinguish two possible alternatives for each case
included in Fig.~\ref{alternatives}, depending on the identity of
the last signal state contained in the previous block. The
probability of this last signal being a coherent state, that we
shall denote as $p$, is calculated in Appendix \ref{ap_A} and it
is given by
\begin{equation}\label{p_bar}
p=\big[p_{succ}+(1-p_{succ})q\big]p_{succ}^{M_{min}}.
\end{equation}
Similarly, $1-p$ represents the probability that the last signal
in a block is a vacuum state. Fig.~\ref{nueva_figura2}
illustrates these two alternatives for the case where Eve sends to
Bob a block of signals containing $M$ coherent states.

Let us now analyze the different scenarios included in
Fig.~\ref{alternatives} in more detail. When Eve sends to Bob a
block of signals containing $M$ coherent states (Case A in
Fig.~\ref{alternatives}) then: If the last signal state of the
previous block is a coherent state, then it turns out that the
average total number of clicks obtained by Bob is given by $Ms$,
where the parameter $s$ has the form
\begin{equation}\label{eqs}
s=1-\exp{(-\mu_\beta)},
\end{equation}
with $\mu_\beta$ being again the mean photon-number of the
coherent states $\ket{\beta{}e^{i\theta_j}}$ sent by Eve.
Otherwise, the average total number of clicks at Bob's side can
be written as $t+(M-1)s$, where the parameter $t$ is given by
\begin{equation}\label{par_t}
t=1-\exp{\big(-\frac{\mu_\beta}{2}\big)}.
\end{equation}
The analysis of the remaining cases is similar. If the first
$m\in(M_{min},M)$ signal states of the block are coherent states,
while the last $M-m$ signals are vacuum states (Case B in
Fig.~\ref{alternatives}) then: If the last state of the previous
block is a coherent state, the average total number of clicks
obtained by Bob is given by $t+ms$. Otherwise, the average total
number of clicks at Bob's side can be written as $2t+(m-1)s$. Eve
can as well send to Bob a block containing first $M-m$ vacuum
states followed by $m\in(M_{min},M)$ coherent states (Case C in
Fig.~\ref{alternatives}). In this situation, if the last state of
the previous block is a coherent state, the average total number
of clicks obtained by Bob is given by $2t+(m-1)s$. Otherwise, the
average total number of clicks has the form $t+(m-1)s$. When Eve
sends to Bob a block of signals where, at least, the first and
the last signals of the block are vacuum states (Case D in
Fig.~\ref{alternatives}) then: If the last state of the previous
block is a coherent state, the average total number of clicks
obtained by Bob is given by $3t+(m-1)s$. Otherwise, the average
total number of clicks has the form $2t+(m-1)s$. The cases E, F,
and G, in Fig.~\ref{alternatives} are completely analogous to the
the cases B, C, and D, respectively. The only difference arises in
the a priori probabilities to be in each of these scenarios. Now,
these a priori probabilities need to be multiplied by the factor
$q$ introduced in Sec.~\ref{seqattack}, {\it i.e.}, by the
probability that Eve actually decides to send $M_{min}$ coherent
states in the block. Finally, when the block contains only vacuum
states (Case H in Fig.~\ref{alternatives}) then: If the last
state of the previous block is a coherent state the average total
number of clicks obtained by Bob is given by $t$. Otherwise, the
average total number of clicks is zero.

After adding all these terms, together with their a priori
probabilities, we obtain that the average total number of clicks
per block of length $M$ at Bob's side in a sequential attack can
be expressed as
\begin{widetext}
\begin{equation}\label{n_clicks}
N_{clicks}^M=pt+p_{succ}^Mu_M+\sum_{M_{min}\leq{}m<M}q^{\delta_{mM_{min}}}(1-p_{succ})p_{succ}^m\Big[v_{m}
+(M-m-1)(1-p_{succ})w_{m}\Big],
\end{equation}
\end{widetext}
where $\delta_{mM_{min}}$ is equal to one if $m=M_{min}$ and it is
zero otherwise, and the parameters $u_M$, $v_{m}$, and $w_{m}$,
are given by
\begin{eqnarray}
u_M&=&(1-2p)t+(M-1+p)s,\nonumber \\
v_{m}&=&(3-2p)t+(2m+p-2)s,\nonumber \\
w_{m}&=&2t+(m-1)s.
\end{eqnarray}

\section{Sequential unambiguous state discrimination
attack}\label{susda}

As already mentioned in the previous section, in this attack Eve
performs unambiguous state discrimination (USD) \cite{usd,chef} of
Alice's signal states. Whenever Eve identifies without error a
signal state sent by Alice then she considers this result as
successful. If the identification process does not succeed, then
she considers it a failure. The probability of obtaining a
successful result per signal state sent by Alice has the form
\cite{usd}
\begin{equation}\label{spusd}
p_{succ}=1-\vert\bra{\alpha}-\alpha\rangle\vert=1-\exp{(-2\mu_\alpha)},
\end{equation}
where $\mu_\alpha$ is the mean photon-number of the signal states
sent by Alice, {\it i.e.}, $\mu_\alpha=\vert\alpha\vert^2$.

Next, we obtain an expression for the quantum bit error rate
(QBER) introduced by Eve with this attack, and also for the
resulting double click rate at Bob's side.

\subsection{Quantum bit error rate}\label{QBER}

The QBER is defined as $N_{errors}/N_{clicks}$, where
$N_{errors}$ represents the average total number of errors
obtained by Bob, and $N_{clicks}$ is again the average total
number of clicks at Bob's side. The parameter $N_{errors}$ can be
expressed as $N_{errors}=(N/M)N_{errors}^M$, with $N_{errors}^M$
denoting the average total number of errors per block of length
$M$. With this notation, the QBER of a sequential attack, that we
shall denote as $Q$, can then be expressed as
\begin{equation}\label{eqqber}
Q=\frac{1}{M}\frac{N_{errors}^M}{G}.
\end{equation}

Next, we obtain an expression for $N_{errors}^M$. We shall
distinguish the same cases as in the previous section, depending
on the number $m$ of coherent states $\ket{\beta{}e^{i\theta_1}},
\ket{\beta{}e^{i\theta_2}},\ldots,\ket{\beta{}e^{i\theta_{m}}}$
inside a block and their position in the block.

Whenever the previous signal of a coherent state inside the block
is a coherent state, then no errors occur since both signals have
the proper relative phase between them. On the contrary, if the
previous signal of a coherent state is a vacuum state or if the
previous signal of a vacuum state is a coherent state then it
turns out that an error can happen with probability
$\exp{(-\mu_\beta/4)}[1-\exp{(-\mu_\beta/4)}]+[1-\exp{(-\mu_\beta/4)}]^2/2
=t/2$, where the parameter $t$ is given by Eq.~(\ref{par_t}). The
error term $[1-\exp{(-\mu_\beta/4)}]^2/2$ that appears in the
previous expression arises from double clicks at Bob's side.
Here, we consider that double click events are not discarded by
Bob, but they contribute to the sifted key. Every time Bob obtains
a double click, he just decides randomly the bit value
\cite{Norbert99}.

Let us begin with Case A in Fig.~\ref{alternatives}. According to
the previous paragraph, if the last signal state of the previous
block is a coherent state, then the average total number of errors
obtained by Bob is zero. Otherwise, it is given by $t/2$. When
the first $m\in{}(M_{min},M)$ signal states of the block are
coherent states (Case B in Fig.~\ref{alternatives}) and the last
state of the previous block is also a coherent state, then the
average total number of errors obtained by Bob is given by $t/2$.
Otherwise, the average total number of errors is $t$. Similarly,
if Eve sends to Bob a block containing first $M-m$ vacuum states
followed by $m\in{}(M_{min},M)$ coherent states (Case C in
Fig.~\ref{alternatives}) and the last signal of the previous
block is a coherent state, then the average total number of errors
is given by $t$. Otherwise, the average total number of errors has
the form $t/2$. Eve can also send a block of signals where, at
least, the first and the last signals of the block are vacuum
states (Case D in Fig.~\ref{alternatives}). Then, if the last
state of the previous block is a coherent state, the average total
number of errors obtained by Bob is given by $3t/2$. Otherwise,
the average total number of errors is $t$. Like in the previous
section, the results for the cases E, F, and G, in
Fig.~\ref{alternatives} can be obtained directly from the cases
B, C, and D, respectively. One only needs to multiply the a
priori probabilities to be in each of these last three scenarios
by the factor $q$. Finally, if the block contains only vacuum
states (Case H in Fig.~\ref{alternatives}) and the last state of
the previous block is a coherent state, then the average total
number of errors is given by $t/2$. Otherwise, the average total
number of errors is zero.

After adding all the terms together, and taking into account the a
priori probabilities of each case, we obtain that the average
total number of errors per block of length $M$ in a sequential USD
attack has the following form
\begin{equation}\label{yqs}
N_{errors}^M=tS,
\end{equation}
where the parameter $S$ is given by
\begin{widetext}
\begin{equation}\label{par_S}
S=\frac{p}{2}+p_{succ}^M\bigg(\frac{1}{2}-p\bigg)
+\sum_{M_{min}\leq{}m<M}q^{\delta_{mM_{min}}}(1-p_{succ})p_{succ}^m\bigg[\bigg(\frac{3}{2}-p\bigg)
+(M-m-1)(1-p_{succ})\bigg].
\end{equation}
\end{widetext}

\subsection{Double click rate}\label{sdcr}

The double click rate at Bob's side, that we shall denote as
$D_c$, is typically defined as $D_c=N_{D_c}/N$, where $N_{D_c}$
refers to the average total number of double clicks obtained by
Bob, and $N$ is again the total number of signal states sent by
Alice. $N_{D_c}$ is given by $N_{D_c}=(N/M)N_{D_c}^M$, with
$N_{D_c}^M$ denoting the average total number of double clicks
per block sent by Eve at Bob's side. The $D_c$ can be written as
\begin{equation}
D_c=\frac{1}{M}N_{D_c}^M.
\end{equation}

In order to obtain an expression for $N_{D_c}^M$, we can again
distinguish the same different cases included in
Fig.~\ref{alternatives}. Double clicks can only occur when the
previous signal of a coherent state is a vacuum state or when the
previous signal of a vacuum state is a coherent state. The
probability to obtain a double click in each of these two
scenarios, that we shall denote as $d$, is given by
\begin{equation}
d=[1-\exp{(-\frac{\mu_\beta}{4})}]^2.
\end{equation}
Otherwise, the probability to have a double click is always zero.
The analysis is then completely equivalent to the one included in
Sec.~\ref{QBER}, one only needs to substitute the parameter $t/2$
by $d$. We obtain, therefore, that the average total number of
double clicks per block sent by Eve in a sequential USD attack can
be written as
\begin{equation}\label{eqdcusd}
N_{D_c}^M=2dS,
\end{equation}
with $S$ given by Eq.~(\ref{par_S}).

\subsection{Evaluation}\label{evalua_usd}

We have seen above that a sequential USD attack can be
parameterized by the block size $M$, the minimum number $M_{min}$
of consecutive successful results within a block that Eve needs
to obtain in order to send Bob a new train of coherent states,
the mean photon-number $\mu_\beta$ of these coherent states sent
by Eve, and the value of the probability $q$, {\it i.e.}, the
probability that Eve actually decides to send $M_{min}$ coherent
states in a block instead of only vacuum states.

\begin{figure}
\begin{center}
\includegraphics[scale=0.37]{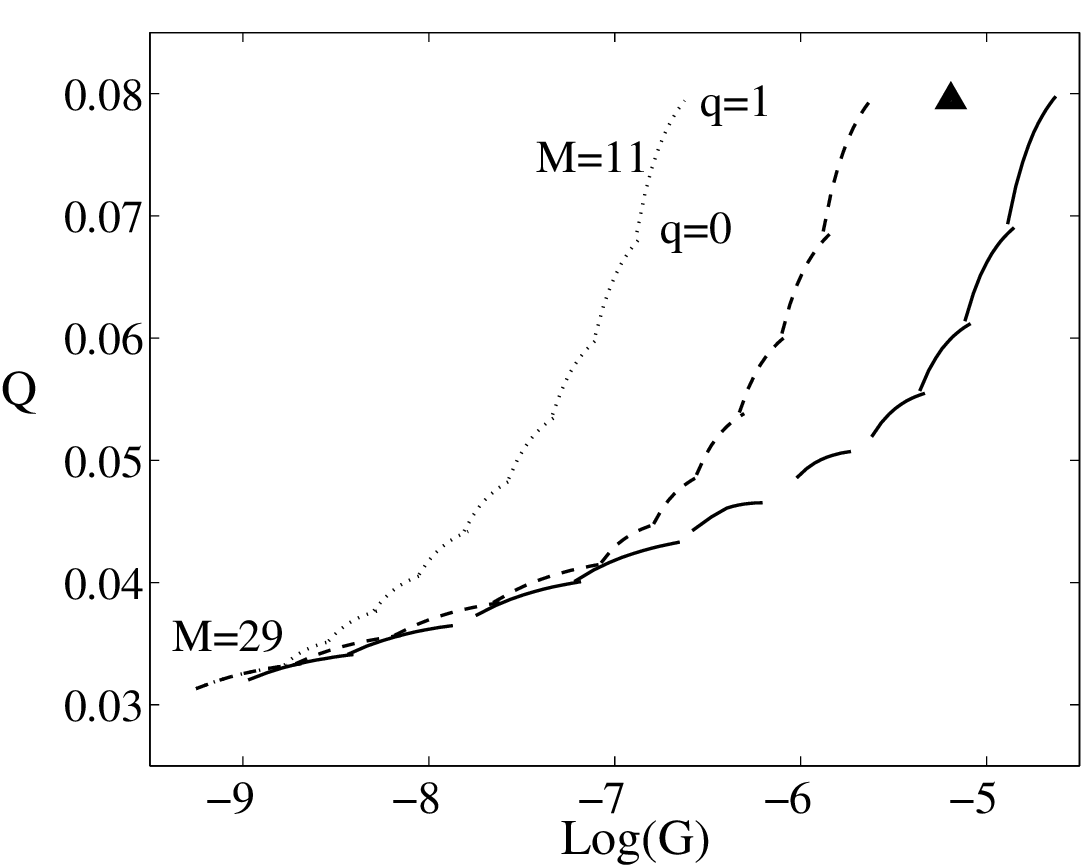}
\end{center}
\caption{Gain versus QBER in a sequential USD attack for
different values of the maximum tolerable double click rate at
Bob's side: $D_c<10^{-8}$ (solid), $D_c<10^{-10}$ (dashed), and
$D_c<10^{-12}$ (dotted). The mean photon number of Alice's signal
states is $\mu_\alpha=0.16$. The triangle represents experimental
data from Ref.~\cite{dpsqkd_exp1}. \label{usd_16}}
\end{figure}

Fig.~\ref{usd_16} shows a graphical representation of the Gain
versus the QBER in this attack for different values of the maximum
tolerable double click rate at Bob's side.
In this example we consider that the mean photon number of Alice's
signal states is given by $\mu_\alpha=0.16$. Moreover, we fix the
value of $M_{min}$ as $M_{min}=\lfloor{}M/2+1\rfloor$ and, for
each given values of the parameters $M$, $q\in[0,1]$, and the
maximum tolerable double click rate obtained by Bob, we perform a
numerical optimization to find the optimal mean photon number
$\mu_\beta$ for each case, {\it i.e.}, the one that provides a
lower QBER for a given value of the Gain. Fig.~\ref{usd_16} also
includes experimental data from Ref.~\cite{dpsqkd_exp1}. According
to these results we find that, unless Alice and Bob reject a
double click rate as low as $10^{-8}$, the DPS QKD experiment
reported in Ref.~\cite{dpsqkd_exp1} would be insecure against a
sequential USD attack. More precisely, our analysis suggest that
in this kind of QKD protocols is not enough for Alice and Bob to
include the effect of the double clicks obtained by Bob in the
QBER \cite{Norbert99}, but it might be very useful for the
legitimate users to monitor also the double click rate to
guarantee security against a sequential attack. The authors of
Ref.~\cite{dpsqkd_exp1} already noticed in
Ref.~\cite{dpsqkd_exp2} that their experiment is not covered by
the existing initial security analysis provided in
Ref.~\cite{dpsqkd2}. Our result is strong as it also shows that
when the double click rate at Bob's side is above $10^{-8}$ no
improved classical communication protocol or improved security
analysis might allow the data of Ref.~\cite{dpsqkd_exp1} to be
turned into secret key.

\begin{figure}
\begin{center}
\includegraphics[scale=0.37]{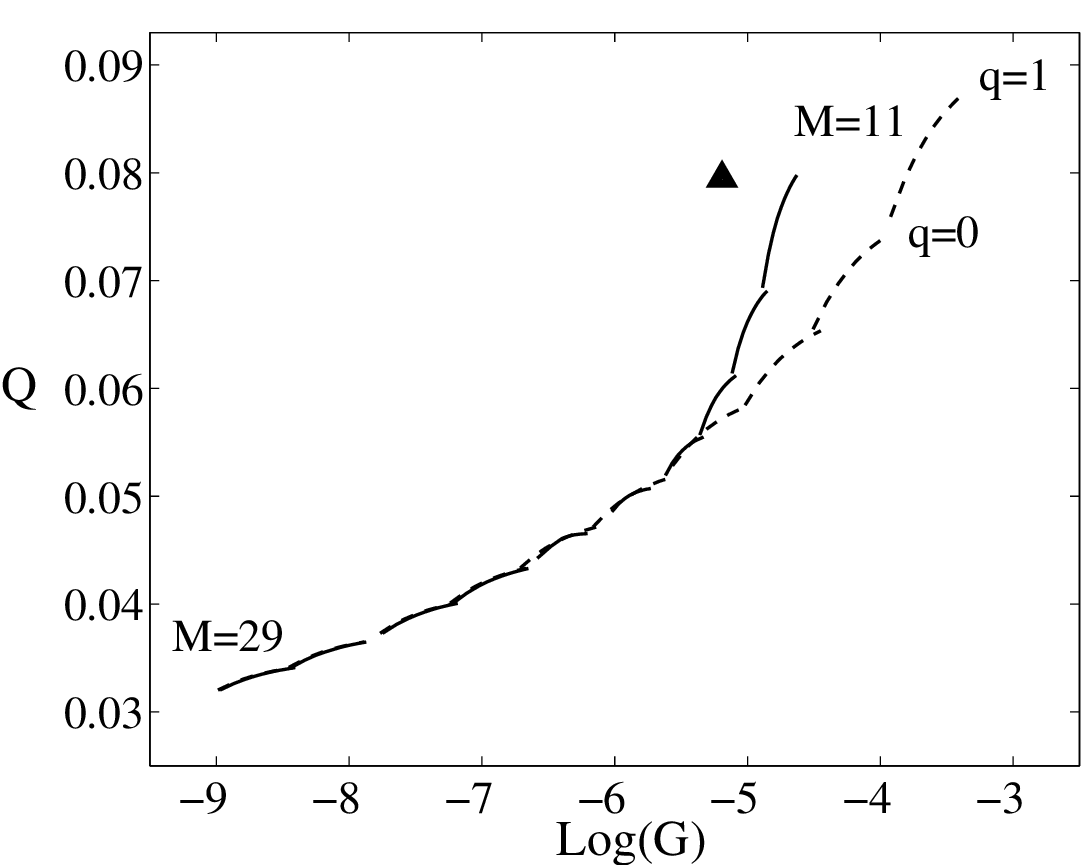}
\end{center}
\caption{Gain versus QBER in a sequential USD attack. The solid
line corresponds to a maximum tolerable double click rate at Bob's
side of $D_c<10^{-8}$. The dashed line represents the case where
Alice and Bob do not monitor separately the double click rate
obtained by Bob. The mean photon number of Alice's signal states
is $\mu_\alpha=0.16$. The triangle represents experimental data
from Ref.~\cite{dpsqkd_exp1}.\label{NDC_016}}
\end{figure}

Fig.~\ref{NDC_016} shows a graphical representation for the case
where Alice and Bob do not monitor separately the double click
rate and Eve can optimize the mean photon number $\mu_\beta$ for
each given values of $M$, $M_{min}=\lfloor{}M/2+1\rfloor$, and the
parameter $q$, without any restriction on the maximum tolerable
double click rate at Bob's side.

\begin{figure}
\begin{center}
\includegraphics[scale=0.37]{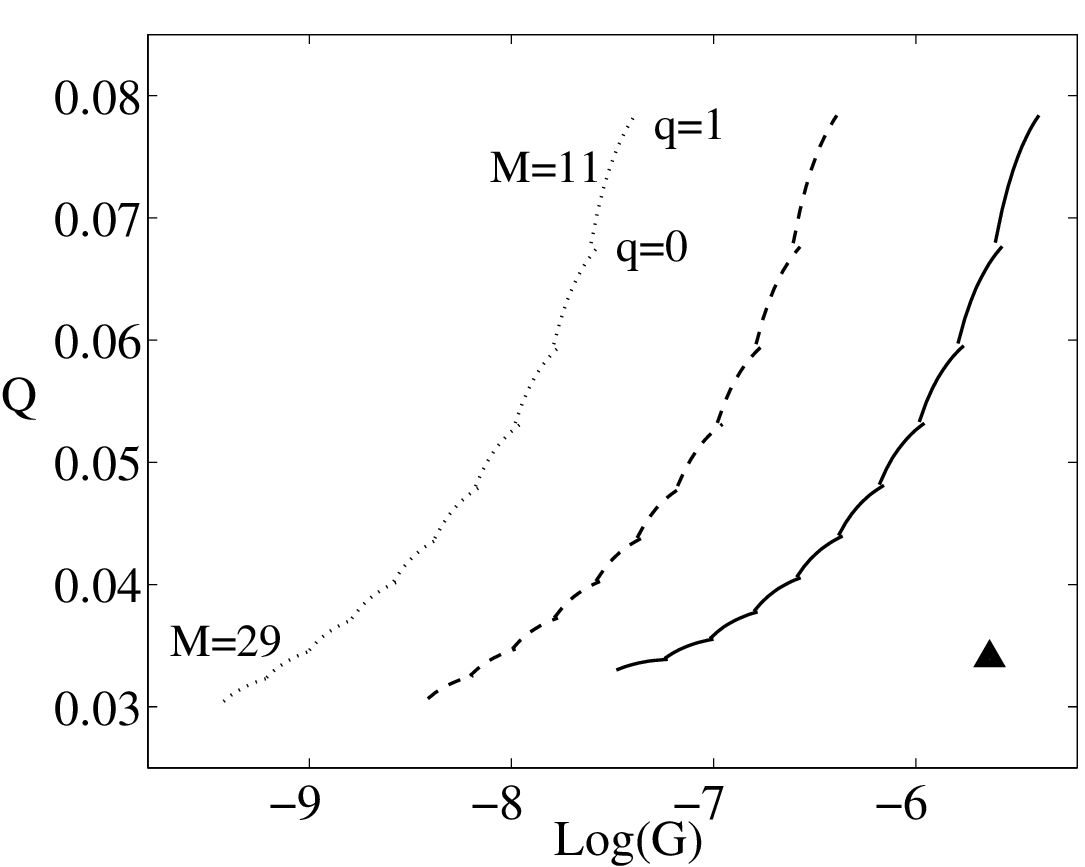}
\end{center}
\caption{Gain versus QBER in a sequential USD attack for
different values of the maximum tolerable double click rate at
Bob's side: $D_c<10^{-10}$ (solid), $D_c<10^{-12}$ (dashed), and
$D_c<10^{-14}$ (dotted). The mean photon number of Alice's signal
states is $\mu_\alpha=0.2$. The triangle represents experimental
data from Ref.~\cite{dpsqkd_exp2}.\label{usd_02}}
\end{figure}

\begin{figure}
\begin{center}
\includegraphics[scale=0.37]{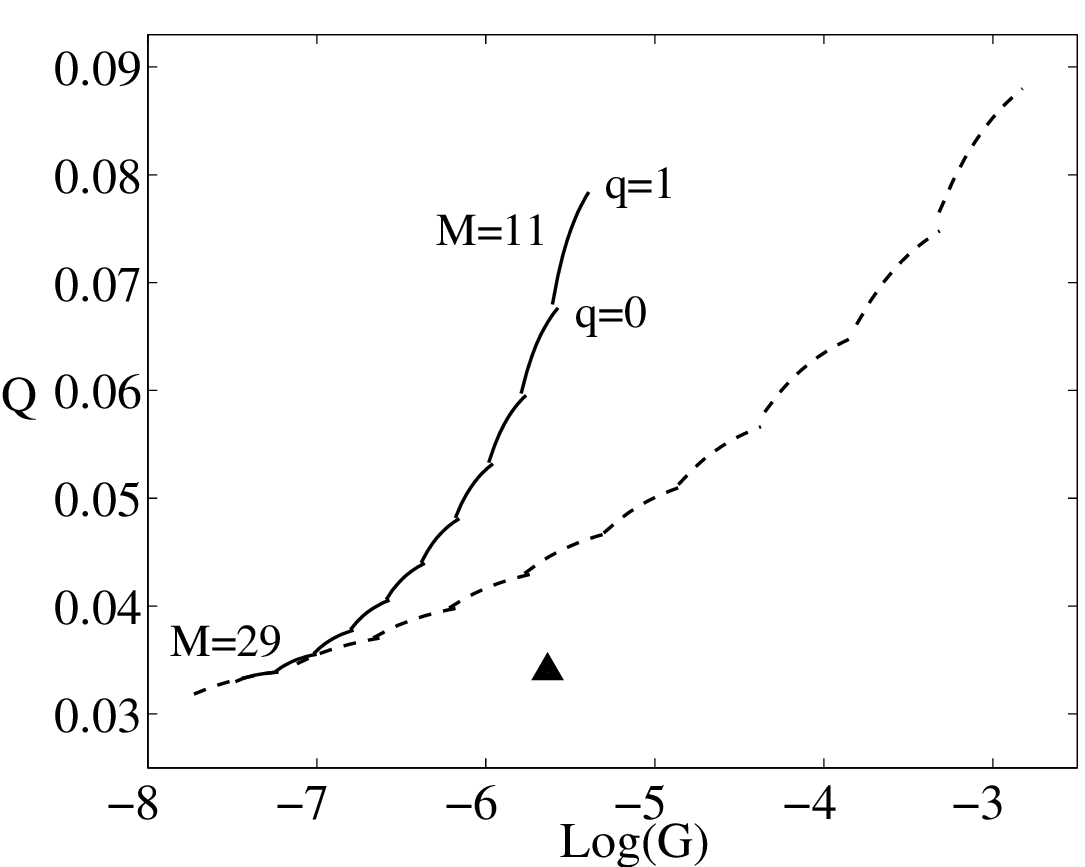}
\end{center}
\caption{Gain versus QBER in a sequential USD attack. The solid
line corresponds to a maximum tolerable double click rate at Bob's
side of $D_c<10^{-10}$. The dashed line represents the case where
Alice and Bob do not monitor separately the double click rate
obtained by Bob. The mean photon number of Alice's signal states
is $\mu_\alpha=0.2$. The triangle represents experimental data
from Ref.~\cite{dpsqkd_exp2}.\label{NDC_02}}
\end{figure}

A similar representation is plotted in Fig.~\ref{usd_02}, but now
for the case $\mu_\alpha=0.2$ and for different values of the
maximum double click rate at Bob's side. In this figure we also
include data from a recent experiment reported in
Ref.~\cite{dpsqkd_exp2}, where the QBER was reduced to a value of
only $3.4\%$.
The scenario where Alice and Bob do not monitor separately the
double click rate obtained by Bob is illustrated in
Fig.~\ref{NDC_02}.
In both cases, our results are consistent with the possibility to
create secret keys.

According to the figures presented in this section, whenever Eve
tries to increase the Gain of this attack by reducing, for
instance, the size $M$ of her blocks, she also increases the
resulting QBER obtained by Bob. The maximum value of the Gain
that Eve can achieve, however, is actually limited by the
probability $p_{succ}=1-\exp{(-2\mu_\alpha)}$ of obtaining a
successful result when distinguishing unambiguously the states
$\ket{\pm\alpha}$. Since, by definition,
$\lfloor{}M/2+1\rfloor\leq{}M_{min}<M$, the minimum value of a
valid block size $M$ is given by $M=3$. This means, in
particular, that in order to maximize the Gain of a sequential USD
attack the best choice for Eve is to select $M=3$ and $M_{min}=2$.
Moreover, we can assume that Eve always sends to Bob $M_{min}$
coherent states $\ket{\beta{}e^{i\theta_1}},
\ket{\beta{}e^{i\theta_2}},\ldots,\ket{\beta{}e^{i\theta_{M_{min}}}}$
when she obtains $M_{min}$ successful results ({\it i.e.},
$q=1$), and that these coherent states have a really high mean
photon number such as she increases Bob's probability of
obtaining a click ({\it i.e.}, $\mu_\beta\gg{}1$ and, therefore,
$s\approx{}1$, $t\approx{}1$, and $d\approx{}1$). Using these
values in Eq.~(\ref{gain}) and Eq.~(\ref{n_clicks}) we obtain
that the maximum value of the Gain in this attack is given by
\begin{equation}\label{maxgain}
G_{max}\approx\frac{1}{3}(6-2p_{succ}-p_{succ}^2)p_{succ}^2.
\end{equation}
In this case the QBER, and the double click rate at Bob's side
are, respectively, given by
$Q\approx(2-p_{succ}-p_{succ}^2)/(6-2p_{succ}-p_{succ}^2)$ and
$D_c\approx{}2(2-p_{succ}-p_{succ}^2)p_{succ}^2/3$.

On the contrary, the minimum value of the Gain occurs when Eve
treats the total number of signals $N$ sent by Alice as a single
block, {\it i.e.}, $M=N$, and she further imposes $M_{min}=M-1$,
$q=0$, and $s\approx{}1$. In this case, the minimum Gain is given
by $p_{succ}^N$, and the QBER and double click rate at Bob's side
are both zero. This scenario corresponds to the situation where
Eve only sends $N$ coherent states $\ket{\beta{}e^{i\theta_1}},
\ket{\beta{}e^{i\theta_2}},\ldots,\ket{\beta{}e^{i\theta_{N}}}$ to
Bob when she succeeds discriminating without error {\it all} the
signal states sent by Alice.

Finally, let us mention that, instead of using an USD measurement
on each signal state sent by Alice, Eve could as well employ the
same detection device like Bob. This sequential attack was very
briefly introduced in Ref.~\cite{dpsqkd2}. In this case, a
successful result is associated with obtaining a click in Eve's
apparatus, while a failure corresponds to the absence of a click.
The train of coherent states $\ket{\beta{}e^{i\theta_1}},
\ket{\beta{}e^{i\theta_2}},\ldots, \ket{\beta{}e^{i\theta_m}}$
that Eve sends to Bob is now selected such as the relative phase
between consecutive signals agree with Eve's measurement results.
If we assume that Eve does not analyze each block of data
independently, but she also includes a proper relative phase
between blocks when the last signal of a previous block and the
first signal of the following one are coherent states, then the
results included in this section also apply to that case.
Otherwise, the QBER in such kind of attack will be always higher
than in a sequential USD attack. However, since Alice's signal
states are typically coherent pulses with small average photon
number ({\it i.e.}, $\vert\alpha\vert^2\ll{}1$), Eve observes
click events only occasionally. In particular, when she uses the
same detection apparatus like Bob then the probability of
obtaining a successful result will be always smaller than the one
of a sequential USD attack. More precisely, this success
probability has now the form $p_{succ}=1-\exp{(-\mu_\alpha)}$,
and is smaller than the success probability given by
Eq.~(\ref{spusd}).
\begin{figure}
\begin{center}
\includegraphics[scale=0.37]{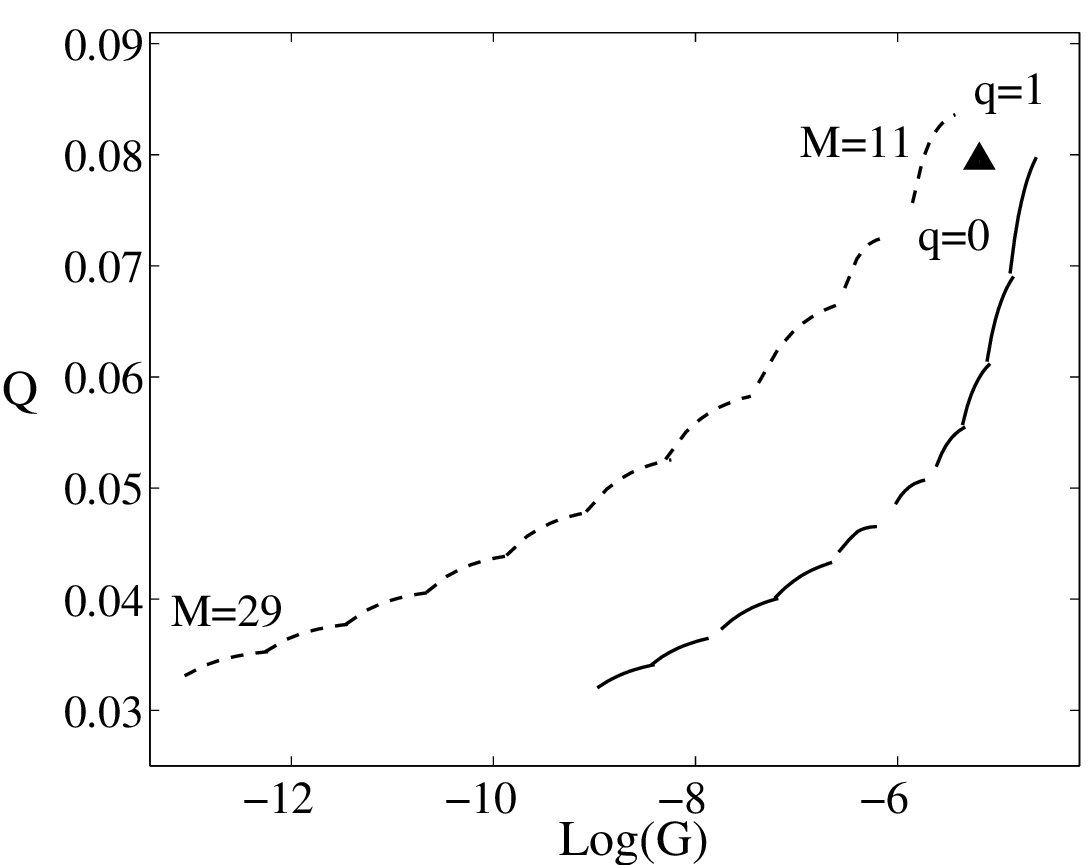}
\end{center}
\caption{Gain versus QBER for a sequential USD attack (solid) and
for a sequential attack where Eve employs the same detection
device like Bob (dashed). The maximum tolerable double click rate
at Bob's side is $D_c<10^{-8}$ and the mean photon number of
Alice's signal states is $\mu_\alpha=0.16$. The triangle
represents experimental data from
Ref.~\cite{dpsqkd_exp1}.\label{comparison}}
\end{figure}
Fig.~\ref{comparison} shows a graphical representation of the Gain
versus the QBER for a sequential USD attack together with a
sequential attack where Eve employs the same detection apparatus
like Bob. In this example the maximum tolerable double click rate
at Bob's side is given by $D_c<10^{-8}$ and the mean photon number
of Alice's signal states is $\mu_\alpha=0.16$. Moreover, we fix
again the value of $M_{min}$ as $M_{min}=\lfloor{}M/2+1\rfloor$
and, for each given values of the parameters $M$ and $q\in[0,1]$,
we perform a numerical optimization to find the optimal
$\mu_\beta$ for each case like before. From the results included
in Fig.~\ref{comparison} we see that a sequential USD attack can
provide tighter upper bounds on the performance of DPS QKD schemes
than a sequential attack with Eve employing the same detection
device like Bob.

\section{Sequential minimum error discrimination attack}\label{medattack}

In this eavesdropping strategy Eve performs first a filtering
operation on each signal state sent by Alice with the intention
to make them, with some finite probability, more
``distinguishable''. Afterwards, Eve measures out each successful
filtered state with a measurement device that gives her the
minimum value of the error probability when identifying the states
\cite{hels,med2}. Her goal is to try to determine whether the
filtered states originate from $\ket{\alpha}$ or from
$\ket{-\alpha}$.

The coherent states sent by Alice can be expressed in some
orthogonal basis $\{\ket{0},\ket{1}\}$ as follows
\begin{equation}
\ket{\pm\alpha}=a\ket{0}\pm{}b\ket{1},
\end{equation}
where we assume, without of generality, that the coefficients $a$
and $b$ are given by
\begin{eqnarray}
a&=&\sqrt{\frac{1}{2}\big[1+\exp{(-2\mu_\alpha)}\big]}\\
b&=&\sqrt{\frac{1}{2}\big[1-\exp{(-2\mu_\alpha)}\big]},\\
\end{eqnarray}
that is, they satisfy, $a\in\mathbb{R}$, $b\in\mathbb{R}$,
$a^2+b^2=1$, and $a>b$ when $\mu_\alpha\neq{}0$.

We shall consider that Eve uses a filtering operation defined by
the following two Kraus operators \cite{Kraus}:

\begin{eqnarray}\label{filter}
A_{succ}(\lambda)&=&\lambda\ket{0}\bra{0}+\ket{1}\bra{1},\\
A_{fail}(\lambda)&=&\sqrt{1-\lambda^2}\ket{0}\bra{0},
\label{filter2}
\end{eqnarray}
where the coefficient $\lambda$ satisfies $\lambda\in[b/a,1]$.
This parameter allows Eve to increase the probability of obtaining
a successful result and, therefore, she can increase the Gain of
her attack. On the other hand, Eve can introduce also more errors
at Bob's side.

Suppose that the filtering operation receives as input the state
$\ket{\pm\alpha}$. The probability of getting a successful result
can be calculated as
$p_{succ}\equiv{}p_{succ}^\lambda=\textrm{Tr}[\ket{\pm\alpha}\bra{\pm\alpha}\
A_{succ}^{\dag}(\lambda)A_{succ}(\lambda)]$. This quantity is
given by
\begin{equation}\label{psuc}
p_{succ}^\lambda=a^2\lambda^2+b^2.
\end{equation}
If the filtering operation succeeded, the resulting normalized
filtered state, that we shall denote as $\ket{\pm\alpha_{succ}}$,
can be calculated as
$\ket{\pm\alpha_{succ}}=(1/\sqrt{p_{succ}^\lambda})\
A_{succ}(\lambda)\ket{\pm\alpha}$. We obtain
\begin{equation}\label{new_sig}
\ket{\pm\alpha_{succ}}=\frac{1}{\sqrt{p_{succ}^\lambda}}
(\lambda{}a\ket{0}\pm{}b\ket{1}).
\end{equation}

As already mentioned previously, in order to decide which signal
state was used by Alice, we consider that Eve follows the
approach of MED. That is, she employs a measurement strategy that
guesses the identity of the signals $\ket{\pm\alpha_{succ}}$ with
the minimum probability of making an error. For the case of two
pure states with equal a priori probabilities, like it is the
case that we have here, the optimal value of the error
probability, that we shall denote as $p_{err}$, is given by
$p_{err}=[1-\sqrt{1-\vert\bra{-\alpha_{succ}}\alpha_{succ}\rangle\vert^2}]/2$
\cite{hels}. From Eq.~(\ref{new_sig}) we obtain, therefore,
\begin{equation}\label{eqerror}
p_{err}=\frac{1}{2}\frac{(a\lambda-b)^2}{a^2\lambda^2+b^2}.
\end{equation}
The von Neumann measurement which can be used to attain this
error probability is given by the optimum detector states
$\ket{\pm}=1/\sqrt{2}(\ket{0}\pm\ket{1})$.

Note that the sequential USD attack introduced in Sec.~\ref{susda}
can then be seen as a special case of this sequential MED attack.
When $\lambda=b/a$, the success probability in a sequential MED
attack is given by
$p_{succ}^{\lambda}=2b^2=1-\exp{(-2\mu_\alpha)}$, which coincides
with the success probability given by Eq.~(\ref{spusd}).
Moreover, in this case the error probability $p_{err}$ is zero.

Next, we obtain an expression for the QBER introduced by Eve with
this attack, and also for the resulting double click rate at
Bob's side.

\subsection{Quantum bit error rate}\label{qbermeda}

From Eq.~(\ref{eqqber}) we learn that in order to obtain an
expression for the QBER in a sequential attack we only need to
find the average total number of errors $N_{errors}^M$ per block
of length $M$.

Now, however, the analysis is slightly different from that
considered in Sec.~\ref{QBER} since two consecutive coherent
states in a block can also produce errors. This arises from the
fact that sometimes Eve does not identify correctly the signal
states $\ket{\pm\alpha}$ sent by Alice. In particular, whenever
the previous signal of a coherent state inside a block is also a
coherent state, then an error can occur with probability
$[p_{err}(1-p_{err})+p_{err}(1-p_{err})]s$, where $p_{err}$ is
given by Eq.~(\ref{eqerror}) and $s$ is given by Eq.~(\ref{eqs}).
This is the probability that only one of the two coherent states
is wrongly identify by Eve and Bob detects the error by means of
a click in his apparatus. We shall denote this error probability
as $\tilde{p}_{err}$. Using Eq.~(\ref{eqerror}), we can write
$\tilde{p}_{err}$ as
\begin{equation}
\tilde{p}_{err}=\frac{1}{2}\bigg(\frac{a^2\lambda^2-b^2}{a^2\lambda^2+b^2}\bigg)^2s.
\end{equation}
If the previous signal of a coherent state is a vacuum state or if
the previous signal of a vacuum state is a coherent state then the
error probability is the same as in Sec.~\ref{QBER}, {\it i.e.},
it has the form $t/2$ with $t$ given by Eq.~(\ref{par_t}).

We can now address the different cases contained in
Fig.~\ref{alternatives} like in the previous sections and obtain
an expression for $N_{errors}^M$ as a function of these two error
probabilities. The analysis is included in Appendix \ref{ap_b}.
We find that $N_{errors}^M$ can be written as
\begin{widetext}
\begin{equation}\label{exqbermeda}
N_{errors}^M=\frac{pt}{2}+p_{succ}^M\tilde{u}_M+\sum_{M_{min}\leq{}m<M}
q^{\delta_{mM_{min}}}(1-p_{succ})p_{succ}^m\Big[\tilde{v}_{m}
+(M-m-1)(1-p_{succ})\tilde{w}_{m}\Big],
\end{equation}
\end{widetext}
where the parameters $\tilde{u}_M$, $\tilde{v}_{m}$, and
$\tilde{w}_{m}$, are given by
\begin{eqnarray}
\tilde{u}_M&=&\frac{(1-2p)t}{2}+(M-1+p)\tilde{p}_{err},\nonumber \\
\tilde{v}_{m}&=&\frac{(3-2p)t}{2}+(2m+p-2)\tilde{p}_{err},\nonumber \\
\tilde{w}_{m}&=&t+(m-1)\tilde{p}_{err},
\end{eqnarray}
and with $p$ given by Eq.~(\ref{p_bar}).

\subsection{Double click rate}

Like in the case of a sequential USD attack, also in this attack
double clicks can happen only when the previous signal of a
coherent state is a vacuum state or when the previous signal of a
vacuum state is a coherent state. The probability to obtain a
double click in each of these two scenarios does not depend on
the value of the phase $\theta_j$ of the coherent state
$\ket{\beta{}e^{i\theta_j}}$ involved, but it depends only on the
mean photon-number $\mu_\beta$. This means that the analysis
included in Sec.~\ref{sdcr} also applies here, and the average
total number of double clicks per block sent by Eve in a
sequential MED attack is also given by Eq.~(\ref{eqdcusd}).

\subsection{Evaluation}

\begin{figure}
\begin{center}
\includegraphics[scale=0.37]{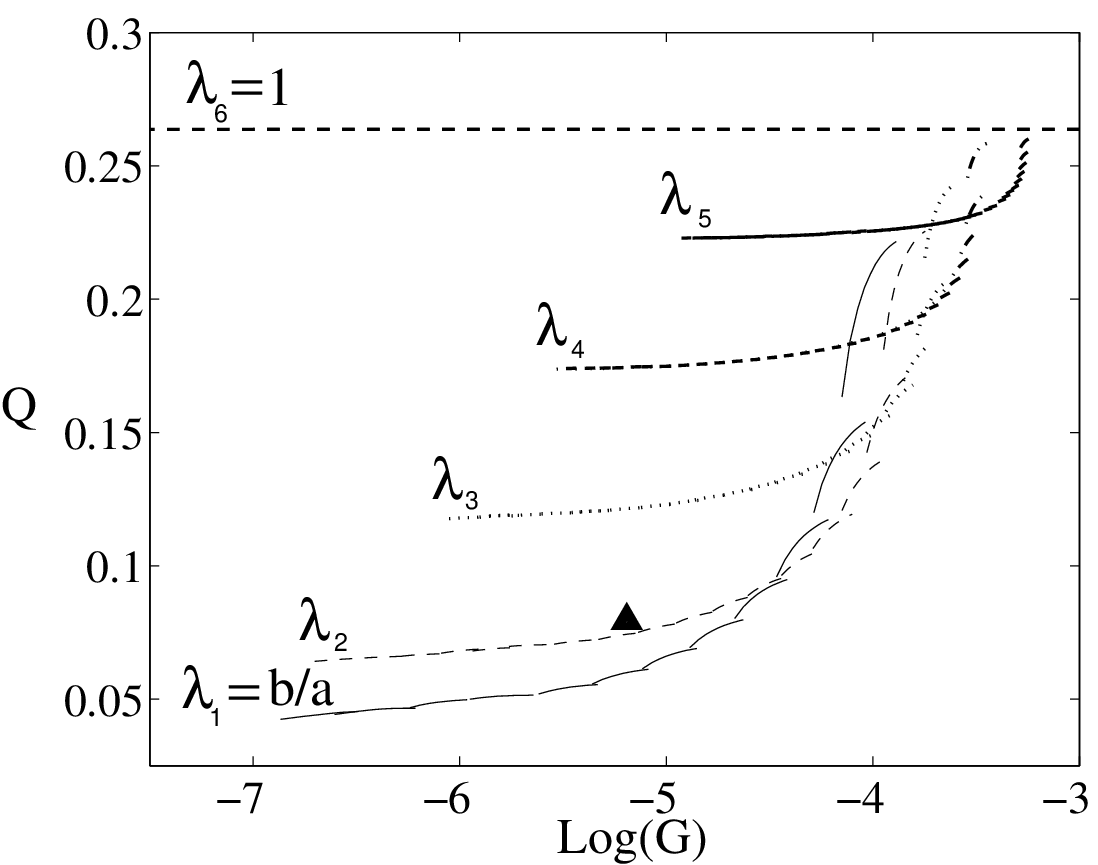}
\end{center}
\caption{Gain versus QBER in a sequential MED attack for a fix
value of the maximum tolerable double click rate at Bob's side
($D_c<10^{-8}$), and for different values of the parameter
$\lambda$: $\lambda_1=b/a$ (solid), $\lambda_2=b/a+(1-b/a)/5$
(dashed), $\lambda_3=b/a+2(1-b/a)/5$ (dotted),
$\lambda_4=b/a+3(1-b/a)/5$ (dashed-dotted),
$\lambda_5=b/a+4(1-b/a)/5$ (thick solid), and $\lambda_6=1$
(thick dashed). The mean photon number of Alice's signal states
is $\mu_\alpha=0.16$. The triangle represents experimental data
from Ref.~\cite{dpsqkd_exp1}. \label{med_016b}}
\end{figure}

In Fig.~\ref{med_016b} we plot the Gain versus the QBER in a
sequential MED attack for a fix value of the maximum tolerable
double click rate at Bob's side ($D_c<10^{-8}$) and for different
values of the parameter $\lambda$. Like in Sec.~\ref{evalua_usd},
we fix the value of $M_{min}$ as $M_{min}=\lfloor{}M/2+1\rfloor$,
and we perform a numerical optimization to find the optimal mean
photon number $\mu_\beta$ for each given values of the parameters
$M$, $q$, and $\lambda$. Moreover, in this example, we consider
that the mean photon number of Alice's signal states is given by
$\mu_\alpha=0.16$ and we also include the experimental data
obtained in Ref.~\cite{dpsqkd_exp1}.

\begin{figure}
\begin{center}
\includegraphics[scale=0.37]{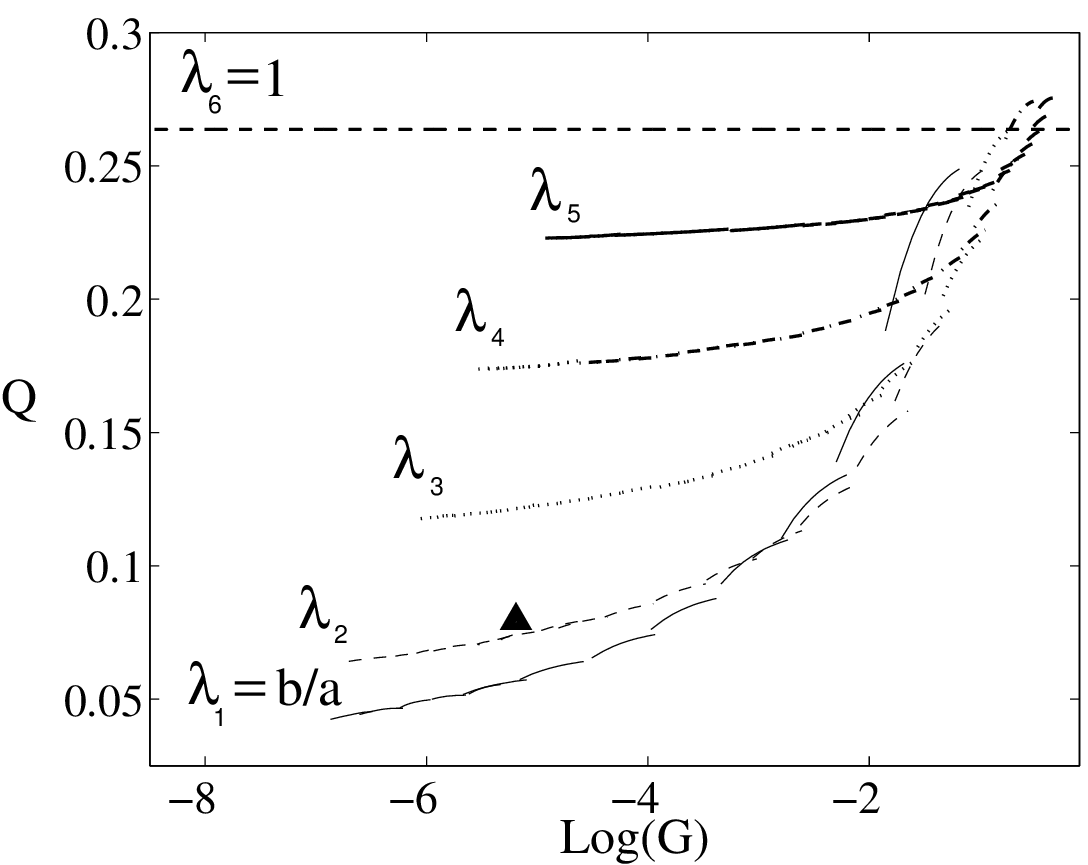}
\end{center}
\caption{Gain versus QBER in a sequential MED attack for the case
where Alice and Bob do not monitor separately the double click
rate obtained by Bob, and for different values of the parameter
$\lambda$: $\lambda_1=b/a$ (solid), $\lambda_2=b/a+(1-b/a)/5$
(dashed), $\lambda_3=b/a+2(1-b/a)/5$ (dotted),
$\lambda_4=b/a+3(1-b/a)/5$ (dashed-dotted),
$\lambda_5=b/a+4(1-b/a)/5$ (thick solid), and $\lambda_6=1$
(thick dashed). The mean photon number of Alice's signal states
is $\mu_\alpha=0.16$. The triangle represents experimental data
from Ref.~\cite{dpsqkd_exp1}. \label{ndc__016b}}
\end{figure}

A similar graphical representation is included in
Fig.~\ref{ndc__016b}, but now for the case where Alice and Bob do
not monitor separately the double click rate and Eve can optimize
the mean photon number $\mu_\beta$ for each given values of $M$,
$M_{min}=\lfloor{}M/2+1\rfloor$, $q$, and the parameter $\lambda$,
without any restriction on the maximum tolerable double click
rate at Bob's side.

While in a sequential USD attack the maximum value of the Gain is
given by Eq.~(\ref{maxgain}), in a sequential MED attack Eve can
always increase the value of the Gain at the expense of also
increasing the resulting QBER at Bob's side, just by incrementing
the parameter $\lambda$. In particular, in the limit case of
$\lambda=1$, {\it i.e.}, the filtering operation is just the
identity operation, we have that $p=1$ and
$p_{succ}^{\lambda=1}=1$. In this situation, the Gain, the QBER,
and the double clock rate at Bob's side are, respectively, given
by $G=1-\exp(-\mu_{\beta})$, $Q=\exp(-4\mu_{\alpha})/2$, and
$D_c=0$. That is, by selecting a proper mean photon number
$\mu_{\beta}$ Eve can always access any high value of the Gain.

\section{CONCLUSION}\label{CONC}

In this paper we have quantitatively analyzed limitations on the
performance of differential-phase-shift (DPS) quantum key
distribution (QKD) protocols based on weak coherent pulses. For
that, we have investigated simple eavesdropping strategies based
on sequential attacks: Eve measures out every coherent state
emitted by Alice and prepares new signal states, depending on the
results obtained, that are given to Bob. Whenever Eve obtains a
predetermined number of consecutive successful measurement
outcomes, then she prepares a train of new coherent pulses that
is forwarded to Bob. Otherwise, Eve sends vacuum signals to Bob
to avoid errors. Sequential attacks transform the original
quantum channel between Alice and Bob into an entanglement
breaking channel and, therefore, they do not allow the
distribution of a secret key.

Specifically, we have considered two possible sequential attacks.
In the first one, Eve realizes unambiguous state discrimination
(USD) of Alice's signal states. When Eve identifies unambiguously
a signal state sent by Alice, then she considers this result as
successful. Otherwise, she considers it a failure. In the second
attack, Eve performs first a filtering operation on each signal
emitted by Alice and, afterwards, she measures out each
successful filtered state following the approach of minimum error
discrimination, {\it i.e.}, she guesses the identity of the
filtered state with the minimum probability of making an error.
As a result, we obtained upper bounds on the maximal distance
achievable by differential-phase-shift quantum key distribution
schemes as a function of the error rate in the sifted key, the
double click rate at Bob's side, and the mean photon-number of
the signals sent by Alice. It states that no key distillation
protocol can provide a secret key from the correlations
established by the users.

Instead of using an USD measurement on each signal state sent by
Alice, like in the first eavesdropping strategy that we
considered, Eve could as well employ the same detection device
like Bob \cite{dpsqkd2}. A successful result is now associated
with obtaining a click in Eve's apparatus, while a failure
corresponds to the absence of a click. However, since Alice's
signal states are typically coherent pulses with small average
photon number, the probability of obtaining a successful result
in this scenario is always smaller than the one of a sequential
USD attack. Therefore, a sequential USD attack can provide
tighter upper bounds on the performance of DPS QKD protocols than
those derived from a sequential attack where Eve uses the same
measurement apparatus like Bob.

While in the standard Bennett-Brassard 1984 (BB84) QKD protocol
with phase randomized weak coherent state sources it generally
suffices that the legitimate users monitor the error rate and
gain of the scheme to guarantee unconditional security, our
analysis suggest that, in DPS QKD, it might be very useful for the
legitimate users to monitor also the double click rate or the
correlations of detection probabilities between adjacent
time-slots. This fact could substantially increase Alice and Bob's
ability in defeating sequential attacks. Therefore, it might be
advantageous for a security proof of DPS QKD to include also
Alice and Bob's knowledge of double click rates and correlations
of detection events. Such a security proof would be rather
different from existing security proofs of the standard BB84
protocol which often involves random permutation and random
sampling arguments.

\section{ACKNOWLEDGEMENTS}

The authors thank Bing Qi for very fruitful discussions on the
topic of this paper.
Financial support from
NSERC, CIPI, CRC program, CFI, OIT, CIAR, PREA, DFG under the
Emmy Noether programme, and the European Commission (Integrated
Project SECOQC) are gratefully acknowledged. This research was
supported by Perimeter Institute for Theoretical Physics.
Research at Perimeter Institute is supported in part by the
Government of Canada through NSERC and by the province of Ontario
through MEDT. M.C. also acknowledges the financial support from a
Post-doctoral grant from the Spanish Ministry of Science (MEC).

\appendix

\section{Probability $p$}\label{ap_A}

In this Appendix we obtain an expression for the probability $p$
that the last signal in a given block is a coherent state
$\ket{\beta{}e^{i\theta_j}}$.

Let $p_m$ be the probability of Eve sending to Bob $m$
consecutive coherent states within a block of length $M$ such
that the last signal of the block is a coherent state. This
probability is given by
\begin{equation}
p_m = \left\{ \begin{array}{ll}
0 & \textrm{if $m<M_{min}$}\\
q(1-p_{succ})p_{succ}^{M_{min}} & \textrm{if $m=M_{min}$}\\
(1-p_{succ})p_{succ}^{m} & \textrm{if $M_{min}<m<M$}\\
p_{succ}^M & \textrm{if $m=M$.}
\end{array} \right.
\end{equation}
For each given block of signals that Eve sends to Bob we have,
therefore, that $p$ can be written as
\begin{equation}\label{p_bar_app}
p=\sum_{m=M_{min}}^M
p_m=\big[p_{succ}+(1-p_{succ})q\big]p_{succ}^{M_{min}}.
\end{equation}
Similarly, $1-p$ represents the probability that the last signal
in a block is a vacuum state.

\section{$N_{errors}^M$ in a sequential minimum error discrimination
attack}\label{ap_b}

In this Appendix we obtain an expression for the average total
number of errors $N_{errors}^M$ per block of length $M$ sent by
Eve in a sequential MED attack.

We shall distinguish the different cases included in
Fig.~\ref{alternatives}, {\it i.e.}, as a function of the number
$m$ of coherent states inside a block and their position in the
block.

Let us begin with Case A in Fig.~\ref{alternatives}. According to
Sec.~\ref{qbermeda}, whenever the last signal state of the
previous block is a coherent state then the average total number
of errors obtained by Bob is given by $M\tilde{p}_{err}$.
Otherwise, it is given by $(M-1)\tilde{p}_{err}+t/2$. If the first
$m\in(M_{min},M)$ signal states of the block are coherent states
(Case B in Fig.~\ref{alternatives}) and the last state of the
previous block is also a coherent state, then the average total
number of errors obtained by Bob is given $m\tilde{p}_{err}+t/2$.
Otherwise, the average total number of errors is
$(m-1)\tilde{p}_{err}+t$. Similarly, if Eve sends to Bob a block
containing first $M-m$ vacuum states followed by
$m\in(M_{min},M)$ coherent states (Case C in
Fig.~\ref{alternatives}) and the last signal of the previous
block is a coherent state, then the average total number of errors
is given by $(m-1)\tilde{p}_{err}+t$. Otherwise, the average total
number of errors has the form $(m-1)\tilde{p}_{err}+t/2$. Eve can
also send to Bob a block of signals where, at least, the first
and the last signals of the block are vacuum states (Case D in
Fig.~\ref{alternatives}). Then, if the last state of the previous
block is a coherent state, the average total number of errors
obtained by Bob is given by $(m-1)\tilde{p}_{err}+3t/2$.
Otherwise, the average total number of errors is
$(m-1)\tilde{p}_{err}+t$.

The results for the cases E, F, and G, in Fig.~\ref{alternatives}
can be obtained directly from the cases B, C, and D,
respectively. One only needs to multiply the a priori
probabilities to be in each of these last three scenarios by the
factor $q$.

Finally, whenever the block that Eve sends to Bob contains only
vacuum states (Case H in Fig.~\ref{alternatives}) and the last
signal of the previous block is a coherent state, then the average
total number of errors is given by $t/2$. Otherwise, the average
total number of clicks is zero.

After including all the a priori probabilities to be in each of
the different cases discussed above, we obtain that the average
total number of errors per block of length $M$ in a sequential MED
attack is given by Eq.~(\ref{exqbermeda}).


\bibliographystyle{apsrev}
\bibliographystyle{apsrev}

\begin{thebibliography}{99}
\bibitem{note1} In order to simplify our notation, from now on we
will employ the term ``coherent state'' only to denote those light
pulses with a mean photon number bigger than zero. A light pulse
with an average photon number equal to zero, although it is also a
coherent state, will be always denoted as a ``vacuum state''.
\end{thebibliography}

\end{document}